\documentclass[aps,prb,twocolumn,showpacs,floatfix,preprintnumbers,amsmath,amssymb]{revtex4-1}
\usepackage{amsmath}
\usepackage{amssymb}
\usepackage{graphicx}
\usepackage{epsfig}
\usepackage{dcolumn}
\usepackage{bm}
\usepackage{array}
\usepackage{xcolor}
\usepackage{multirow}
\usepackage{booktabs}

\begin{document}
\title{Higher-order spacing ratios in random matrix theory and complex quantum systems}
\author{S. Harshini Tekur}
\email{E-mail: harshini.t@gmail.com}
\author{Udaysinh T. Bhosale}
\email{E-mail: udaybhosale0786@gmail.com}
\author{M. S. Santhanam}
\email{E-mail: santh@iiserpune.ac.in}
\affiliation{Indian Institute of Science Education and Research, 
Dr. Homi Bhabha Road, Pune 411 008, India}

\begin{abstract}
The distribution of the ratios of nearest neighbor level spacings has become a popular
indicator of spectral fluctuations in complex quantum systems such as the localized and 
thermal phases of interacting many-body systems, quantum chaotic systems, and in atomic
and nuclear physics. In contrast to the level spacing  distribution, which requires
the cumbersome and at times ambiguous unfolding procedure, the ratios of spacings do
not require unfolding and are easier to compute. In this work, for the class of 
Wigner-Dyson random matrices with nearest neighbor spacing ratios $r$ distributed
as $P_{\beta}(r)$ for the three ensembles indexed by $\beta=1,2, 4$, their $k-$th order
spacing ratio distributions are shown to be identical to $P_{\beta'}(r)$, where $\beta'$,
an  integer, is a function of $\beta$ and $k$. This result is shown for Gaussian and 
circular ensembles of random matrix theory and for several physical systems such
as spin chains, chaotic billiards, Floquet systems and measured nuclear resonances.

\end{abstract}

\maketitle

\section{Introduction}

Spectral fluctuations in complex quantum systems contain information about their
physical character.
In tight binding models and crystalline lattices that display metal to insulator transition, the
metallic and insulator regimes are distinguishable based on their spectral fluctuations\cite{rmt-mit}.
In interacting spin chains, many body localized and thermal phases also carry distinct
spectral signatures\cite{nand}.
In quantum systems with chaotic classical limit\cite{reichl}, the fluctuations indicate if
the system is integrable, chaotic or a mixture of both these types of dynamics\cite{bgs}.

Of the spectral fluctuation measures modeled through random matrix theory (RMT)\cite{mehta2004}, the most popular
one is the nearest neighbor level spacings, $s_i=E_{i+1}-E_i$, where $E_i, i=1,2, \dots$
are the discrete eigenvalues of a quantum operator. For time-reversal invariant systems
without spin degree of freedom, RMT predicts that the spacings are
Wigner distributed\cite{wigner}, $P(s) = (\pi/2) s \exp(-\pi s^2/4)$ indicating the
presence of level repulsion.
In physical systems, Wigner distribution is associated with the metallic regime
of tight binding models \cite{mbl1}, thermal phase of many-body systems\cite{mbl2}, chaotic quantum 
systems\cite{stockmann} such as coupled oscillators\cite{qo} and atoms in strong external 
fields\cite{hatom} and many others\cite{misc}.
This is in contrast to the class of integrable systems
such as many body localized phase of interacting systems\cite{hosr},
which display level clustering through Poissonian spacing distribution, $P(s) = \exp(-s)$.

In practice, spacing distributions have to be computed after removing the system
dependent spectral features, {\it i.e.}, the average part of the density of states, through
a cumbersome and non-unique numerical procedure of unfolding the spectra.
Further, in many-body systems (for example, Bose-Hubbard model) at large interaction 
strengths, the density of states is not a smooth function of energy and hence
separating it into an average and fluctuating part, a pre-requisite for
spectral unfolding, becomes non-trivial \cite{huse,eBH}.

These difficulties have been overcome through the use of
spacing ratio\cite{huse} $r_i = s_{i+1}/s_i, i=1,2,\dots$, since it does
not depend on the local density of states and consequently does not require unfolding.
The RMT averages for the spacing ratios, drawn from three standard random matrix
ensembles with co-dimension $\beta=1,2$ and 4 corresponding to the Gaussian orthogonal ensemble (GOE), 
unitary ensemble (GUE) and symplectic ensemble (GSE), have been obtained as \cite{atas1,atas2},
\begin{equation}
P(r,\beta) = C_{\beta} ~ \frac{(r + r^2)^\beta}{(1+r+r^2)^{1+\frac{3}{2}\beta}},
\label{eq1}
\end{equation}
where $C_\beta$ is a constant that depends on $\beta$.

Several variants of spacing ratios have been studied recently in different contexts\cite{ratios}.
However, few results exist for the distribution of non-overlapping higher-order 
spacing ratio \cite{nod} defined by
\begin{align}
 r_i^{(k)} = \frac{s_{i+k}^{(k)}}{s_{i}^{(k)}} = \frac{E_{i+2k}-E_{i+k}}{E_{i+k}-E_i}, ~~~~~~~
i,k=1,2,3,\dots .
\label{hosr}
\end{align}
Nearest neighbor spacing ratio $r$ probes fluctuations in spectral scales
of the order of unit mean spacing, whereas $r^{(k)}$ probes fluctuations in spectral interval
of $k$ mean spacings.  In many physical situations, knowledge of
spectral fluctuations at larger spectral intervals is useful.
For quantum chaotic systems with a classical limit,
semiclassical theories \cite{semiclass} dictate that the higher-order spectral fluctuations would
be related to short time periodic orbits, effectively acting as a probe of short time
dynamics\cite{stockmann}, at shorter than Heisenberg time-scale. The rare-region effects or 
Griffith effects \cite{griffith} in the vicinity of many-body localization transition influences
the transport and entanglement properties, whose time-scales can be probed by the 
higher-order spacing ratios.

\section{Distribution of higher-order spacing ratios}
Thus, higher-order spacing ratios have profound physical implications apart from being an
intrinsic object of interest in random matrix theory \cite{guhr, rmt-mbl, forrester}. In this work, for the random 
matrices, of order $N>>1$, belonging to the Gaussian and circular ensembles of random matrix theory,
we present compelling numerical evidence to demonstrate an elegant relation between
non-overlapping $k-$th order spacing ratio distribution $P^k(r,\beta)$ and the nearest neighbor
spacing ratio distribution $P(r,\beta')$:
\begin{align}
P^k(r,\beta) & = P(r,\beta'), ~~~~~\beta=1,2,4, \label{eq2}\\
\beta' & =  \frac{k(k+1)}{2}\beta+(k-1),  ~~~~k \geq 1.
\label{eq3}
\end{align}
This scaling relation is the central result of this paper. Note that $4 \le \beta'< \infty$ 
can take large integer values and, unlike $\beta=1,2,4$, does not have corresponding 
random matrix model as yet. Thus, Eq. \ref{eq2} may be considered as a generalization
of the Wigner surmise, that holds good for integer values of $\beta>0$. For $0\leq\beta\leq1$,
a result that can be considered a special case of the scaling relation Eqs. \ref{eq2}-\ref{eq3}
has been proved for generalized $\beta$ ensembles at the level of the joint probability distribution
of eigenvalues\cite{forrester_beta}. It is also pertinent to point out that similar relation
between the higher-order and nearest neighbor {\sl spacing distributions} had been proposed earlier
without rigorous proof \cite{porter,abulmagd}, though their validity had never been tested on spectra
from random matrices or physical systems. One exception is the well-known relation that the
next-nearest neighbor $(k=2)$ level spacings of levels from circular orthogonal ensemble are
distributed as the nearest neighbor $(k=1)$ spacings of levels from circular symplectic ensemble
\cite{dysonmehta}. In the limit of large matrix dimensions, this is known to be valid for the
corresponding Gaussian ensemble as well.

Remarkably, the functional form of $P^k(r,\beta)$ is identical to $P(r,\beta')$ with order
of the spacing ratio $k$ and Dyson index $\beta$ dependence entering through the modified 
parameter $\beta'$. In the rest of the paper, numerical evidence from random matrices and 
from physical systems such as spin chain, quantum billiards and measured nuclear resonances 
are presented, apart from some well-studied models of quantum chaos like the kicked top and 
the intermediate map. Further, Eq. \ref{eq2} suffers from strong finite size effects and we 
discuss disparate cases that have different rates of convergence to Eq. \ref{eq3}.

\section{Results}

\subsection{Random matrix spectra} 

\begin{figure}
\includegraphics*[width=3.6in]{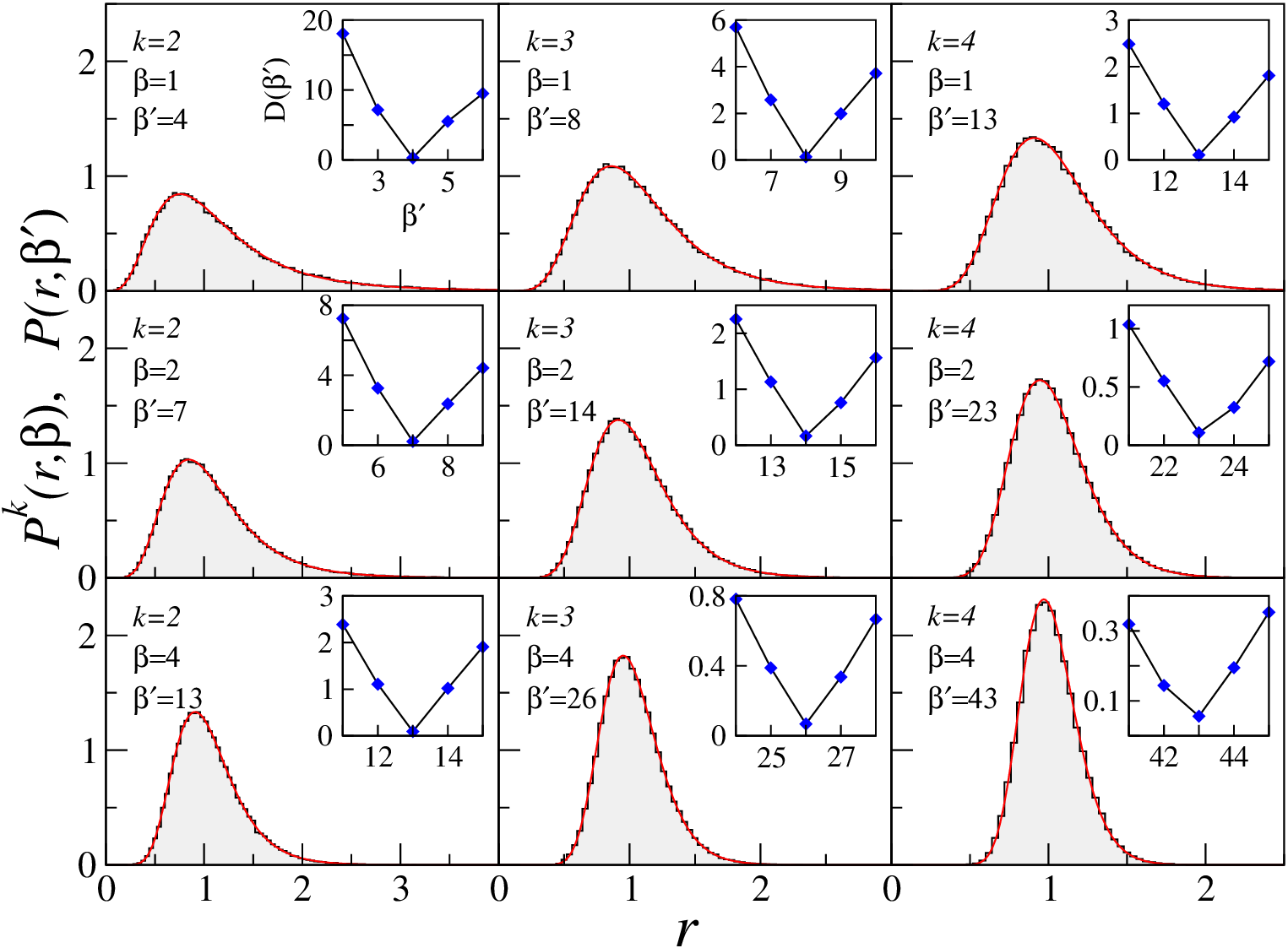}
\caption{Distribution of $k$-th order spacing ratios (histograms) for the spectra of random
matrices drawn from GOE, GUE and GSE and the distribution $P(r,\beta')$ (solid line) with
$\beta'$ given by Eq. \ref{eq3}. (Inset) shows $D$ as a function of $\beta'$.
}
\label{fig1}
\end{figure}

First, Eqs. \ref{eq2}-\ref{eq3} are verified for the spectra 
computed from random matrix ensembles.
The eigenvalues of random matrices (drawn from Gaussian ensembles) of order $N=10^5$ are
computed for $\beta=1,2$ and 4. The resulting histograms of higher-order spacing ratios shown in
Fig. \ref{fig1} are averaged over 1000 realizations.
The solid curves in this figure represent $P(r,\beta')$ and its excellent agreement with
the histograms points to the validity of Eq. \ref{eq2}. Corresponding averages $\langle r\rangle_{th}$, as calculated
theoretically from Eq. \ref{eq2}, and determined from numerics $\langle r\rangle_{G}$ are shown in Table \ref{table1}.
The scaling in Eqs. \ref{eq2}-\ref{eq3} holds good
for the circular ensembles of random matrix theory as well. An excellent agreement with
the postulated scaling relation is observed (not shown here) for the circular orthogonal, unitary and symplectic ensembles
(COE, CUE, and CSE respectively) as well, and the corresponding results for averages $\langle r\rangle_{C}$
are displayed in Table \ref{table1}. 

\begin{table}
\centering
\setlength{\tabcolsep}{10pt}
\renewcommand{\arraystretch}{1.2}
\begin{tabular}{|l|l|l|l|l|l|}

\hline
$\beta$                		& k & $\beta'$ & $\langle r\rangle_{th}$       &  $\langle r\rangle_{G}$ &  $\langle r\rangle_{C}$  \\ [10pt]\hline
\multirow{3}{*}[-1.1em]{1}     & 2 & 4  & 1.1747 & 1.1757 & 1.1767 \\ [10pt]
                               & 3 & 8  & 1.0855 & 1.0847 & 1.0860 \\ [10pt]
                               & 4 & 13 & 1.0521 & 1.0518 & 1.0524 \\ [10pt]\hline
\multirow{3}{*}[-1.1em]{2}     & 2 & 7  & 1.0980 & 1.0976 & 1.0969 \\ [10pt]
				& 3 & 14 & 1.0483 & 1.0478 & 1.0478 \\ [10pt]
				& 4 & 23 & 1.0293 & 1.0289 & 1.0291 \\ [10pt]\hline
\multirow{3}{*}[-1.1em]{4} 	& 2 & 13 & 1.0521 & 1.0522 & 1.0525 \\ [10pt]
				& 3 & 26 & 1.0259 & 1.0258 & 1.0262 \\ [10pt]
				& 4 & 43 & 1.0156 & 1.0156 & 1.0158 \\ \hline
\end{tabular}
\caption{The average value of $r$, as calculated theoretically from Eq. \ref{eq2} ($\langle r\rangle_{th}$)  and as determined
numerically from data for Gaussian ($\langle r\rangle_{G}$) and circular ensembles ($\langle r\rangle_{C}$)
is shown for different values of $k$ and $\beta'$. }
\label{table1}
\end{table}

Further, to quantitatively check that the value of $\beta'$ predicted by Eq. \ref{eq3} is precisely the one that 
best fits the histogram $P^k(r,\beta)$ obtained from random matrix simulations, 
we compute the difference between the cumulative distributions defined as,
\begin{align}
D(\beta')=\sum_i \left| I^k(r,\beta) - I(r_i,\beta') \right|,
\end{align}
where $I(r,\beta')$ and $I^k(r,\beta)$ are the cumulative distributions corresponding, respectively, 
to $P(r,\beta')$ and $P^k(r,\beta)$. Then, the value of $\beta'$ for which $D(\beta')$ is minimum is 
the one that best fits the observed histogram. The insets in Fig. \ref{fig1} display the quantitative 
verification of scaling in Eqs. \ref{eq2}-\ref{eq3}. As seen in the insets, the minima of $D(\beta')$
remarkably coincides with the value of $\beta'$ predicted by Eq. \ref{eq3}. As an additional verification,
the Kolmogorov-Smirnov test\cite{kstest} has also been performed for this data, and the $p$-values obtained
indicate that the histograms correspond to the predicted distribution in Eq. \ref{eq2} with a very 
high probability.

\subsection{Many-body systems (GOE class)}

\begin{figure}
\includegraphics*[width=3.4in]{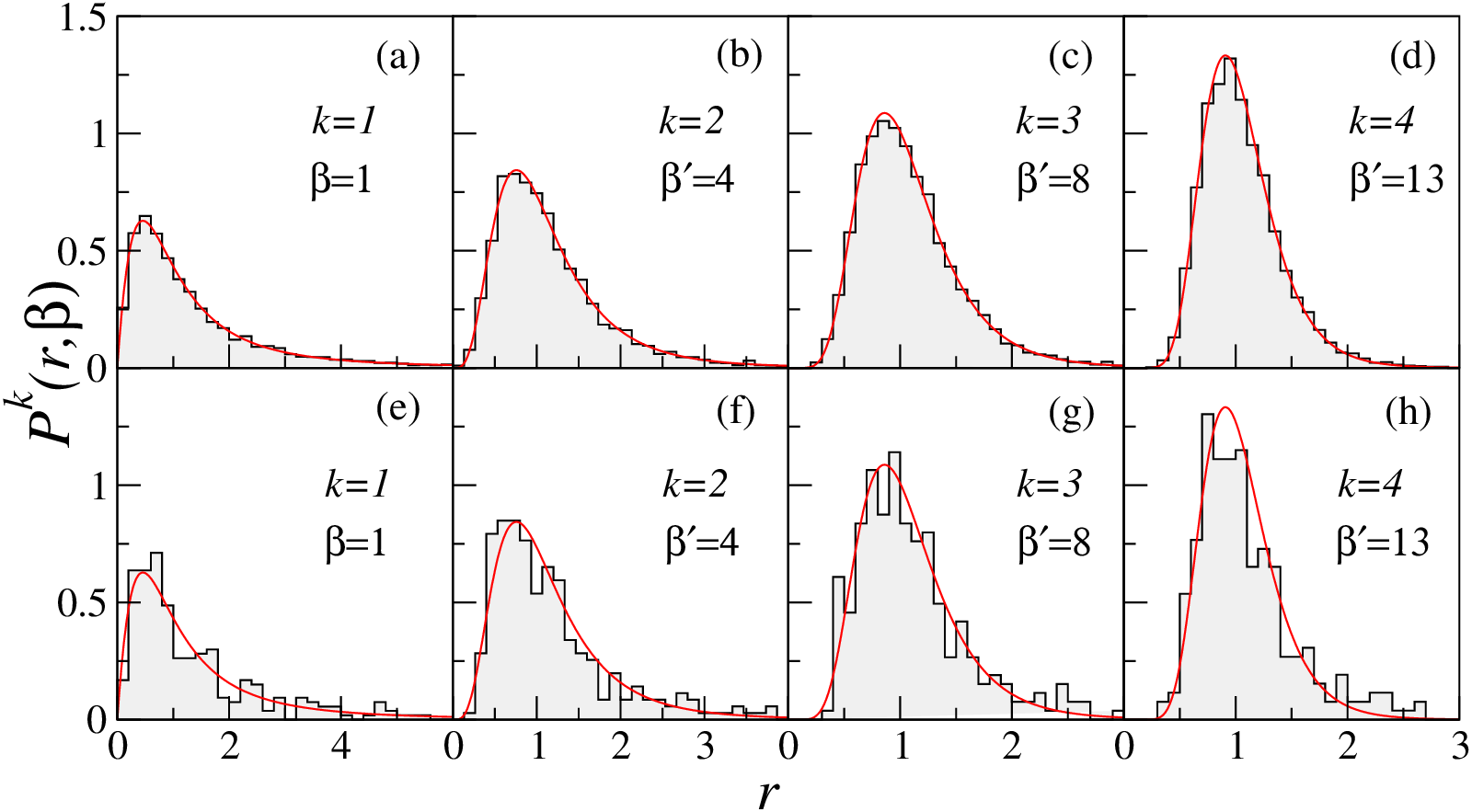}
\caption{Distribution of $k$-th spacing ratio for many-body systems 
of the GOE class ($\beta=1$). The histograms are for the computed spectra 
from a disordered spin chain (upper panel) and nuclear resonance of $^{167}Er$ atom (lower panel).
The solid line corresponds to $P(r,\beta')$ predicted by Eqs. \ref{eq2}-\ref{eq3}, 
with $\beta'$=1, 4, 8 and 13 for $k$=1 to 4.}
\label{goesystem}
\end{figure}

Figure \ref{goesystem} shows validity of the 
scaling relation (Eqs. \ref{eq2}-\ref{eq3}) for
two many-body systems, namely, {\it (i)} a one-dimensional disordered spin-1/2 chain, 
and {\it (ii)} for experimentally measured nuclear resonances of the 
Erbium atom. Both are examples of many-body systems whose nearest-neighbor spectral statistics
had been well-established as coinciding with that of GOE\cite{spinchains,nuclear}. The eigenvalues for the spin chain are 
obtained by diagonalizing the Hamiltonian\cite{santos}
\begin{align}
 H&=\sum_{i=1}^L \omega S_i^z + \epsilon_d S_d^z \nonumber \\
   &+\sum_{i=1}^{L-1}[J_{xy}(S_i^xS_{i+1}^x+S_i^yS_{i+1}^y)+J_zS_i^zS_{i+1}^z].
\label{schain1}
\end{align}
Here, $L$ is the length of the chain and $S^{x,y,z}_i$ are the spin operators in three directions, acting on site $i$. The first 
term of the Hamiltonian represents a static magnetic field in the $z$-direction, accounting for 
a Zeeman splitting of strength $\omega$ at all sites, except the defect site $d$ where it is 
$\epsilon_d+\omega$. The second term, by itself is the well-known XXZ Hamiltonian, and
couples nearest-neighbor spins in all directions, with $J_{xy}$ (taken here to be $1$) being coupling strength along 
$x$- and $y$- directions, and $J_z$ (taken as $0.5$) that along the $z$-direction. For the spectra from the
Hamiltonian in Eq. \ref{schain1}, the upper panel of Fig. \ref{goesystem}
displays a good agreement between the computed $k$-th spacing ratio distribution and
$P(r,\beta')$ given by Eqs. \ref{eq2}-\ref{eq3}. Finite size effects have been discussed for this system
in Fig. \ref{convergence}(c), by varying $L$, which changes the Hilbert space dimension.
For the distributions shown in Figs. \ref{goesystem}(a-d), the length of the spin chain was considered to be
$L=14$, the site of the disorder was taken to be at $L/2$, and the magnitude of the disorder was $\epsilon_d=0.5$.
A similar excellent agreement can be inferred from the lower panel of Fig. \ref{goesystem} for the
experimentally measured data for neutron resonances of the Erbium atom \cite{eratom}. 
Even with only about 200 measured resonances, a good agreement with the theoretical form of $P^k(r,\beta')$
is observed for $k=1$ to 4. 

\subsection{Disordered spin chain and quantum chaotic system (GUE class)}

\begin{figure}
\includegraphics*[width=3.4in]{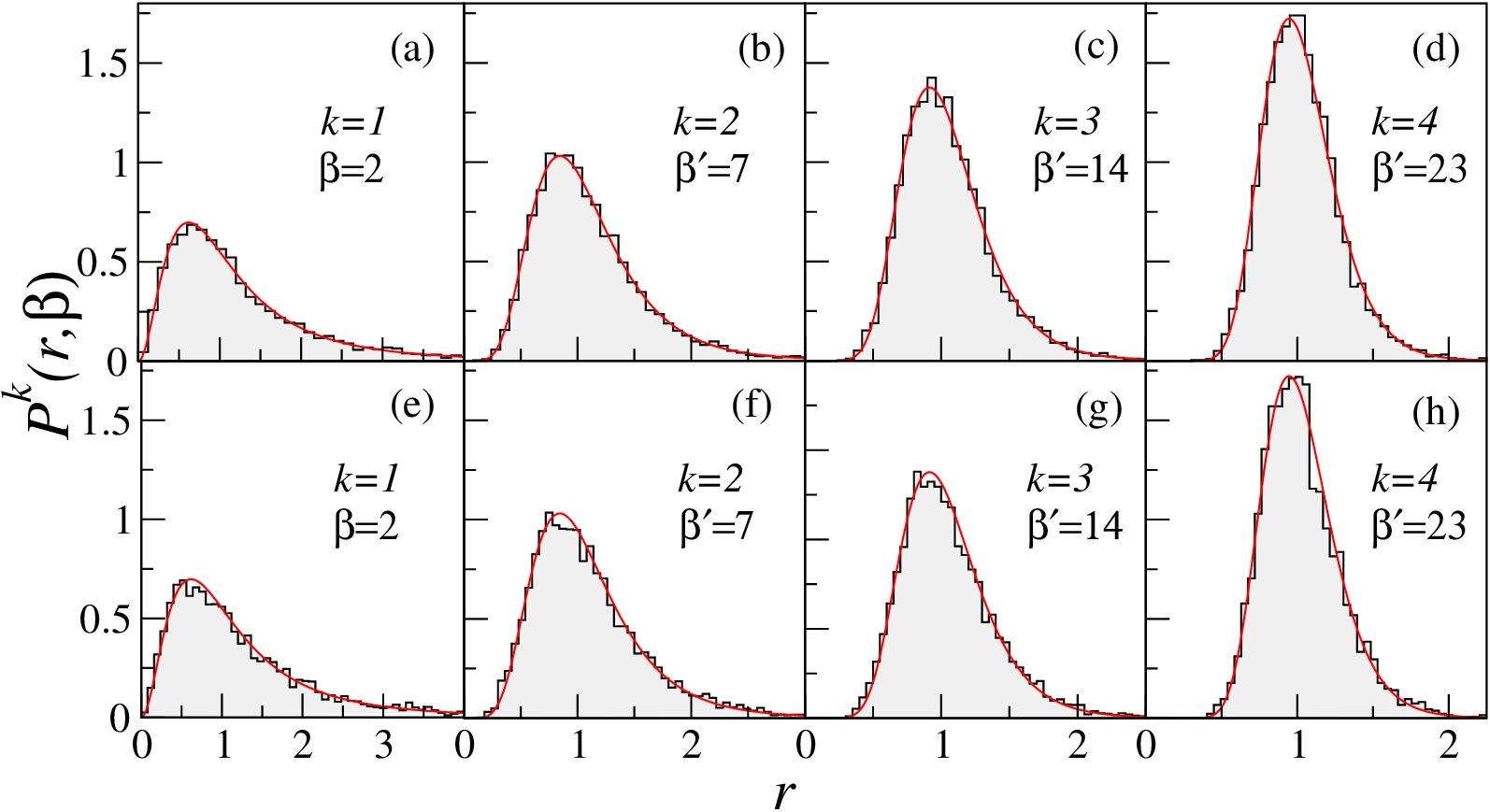}
\caption{Distribution of $k$-th spacing ratio for physical systems of the GUE 
class ($\beta=2$). Histogram is for a  spin chain with a three-spin interaction (upper panel),
and chaotic billiards with a magnetized ferrite strip (lower panel). The solid line represents
the predicted $P(r,\beta')$, with $\beta'=$2, 7, 14 and 23 for $k$=1 to 4.}
\label{guesystem}
\end{figure}

The validity of Eqs. \ref{eq2}-\ref{eq3} for two physical systems belonging to GUE symmetry class
is discussed. They are, (i) a one-dimensional disordered spin-1/2 system \cite{guespinchain}, 
and (ii) a quantum billiard without time reversal symmetry \cite{stodietz}. They are not invariant under time reversal symmetry
and hence belong to the GUE class.
The Hamiltonian for the disordered spin chain is 
\begin{align}
 H=\sum_{i=1}^L[J_1(\textbf{S}_i\cdot\textbf{S}_{i+1}) + h_iS_i^z + 
        J_2(\textbf{S}_i\cdot(\textbf{S}_{i+1}\times\textbf{S}_{i+2})],\nonumber
\end{align}
in which $J_1$ and $J_2$ represent strength of coupling between sites.
The first term (by itself, the Heisenberg spin chain) corresponds to nearest neighbor 
couplings in all directions, with $J_1$ giving the strength of the coupling. The second
term introduces a Gaussian distributed, random magnetic field of mean 0 and strength $h_i$ in 
the $z$-direction. The third term breaks time reversal symmetry by introducing a three-spin interaction
with the nearest as well as the next-nearest neighbor couplings with strength $J_2$. The parameters used 
to obtain data for Figs. \ref{guesystem}(a-d) are $L=12$, $h/J_1=1$ and $J_2/J_1=1$, 
with open boundary conditions. The computed spacing ratio distribution $P^k(r_i,\beta=2)$ shown in
the upper panel of Fig. \ref{guesystem} for $k=2,3,4$ is consistent with $P(r,\beta')$.

An experimentally realized quantum chaotic billiard \cite{guebilliards} has been simulated here,
which is a microwave cavity (of size $\sim17$ inches) placed in a static magnetic field, in which
time reversal symmetry is broken due to the presence of a magnetizd ferrite strip placed on one of
the walls. The simulation uses exactly the same configuration and parameters described in Ref. \cite{guebilliards}.
The eigenvalues were computed by solving the Helmholtz equation with suitable boundary conditions, 
and its fluctuations coincide with GUE statistics.
As seen in lower panel of Fig. \ref{guesystem}, the distribution of $k$-th spacing ratios 
provides another instance of the validity of the scaling relation in
Eqs. \ref{eq2}-\ref{eq3} for GUE systems with $\beta$=2.

\subsection{Spectra of Floquet systems}
\begin{figure}[t]
\vspace{0.5cm}
\includegraphics*[width=3.4in]{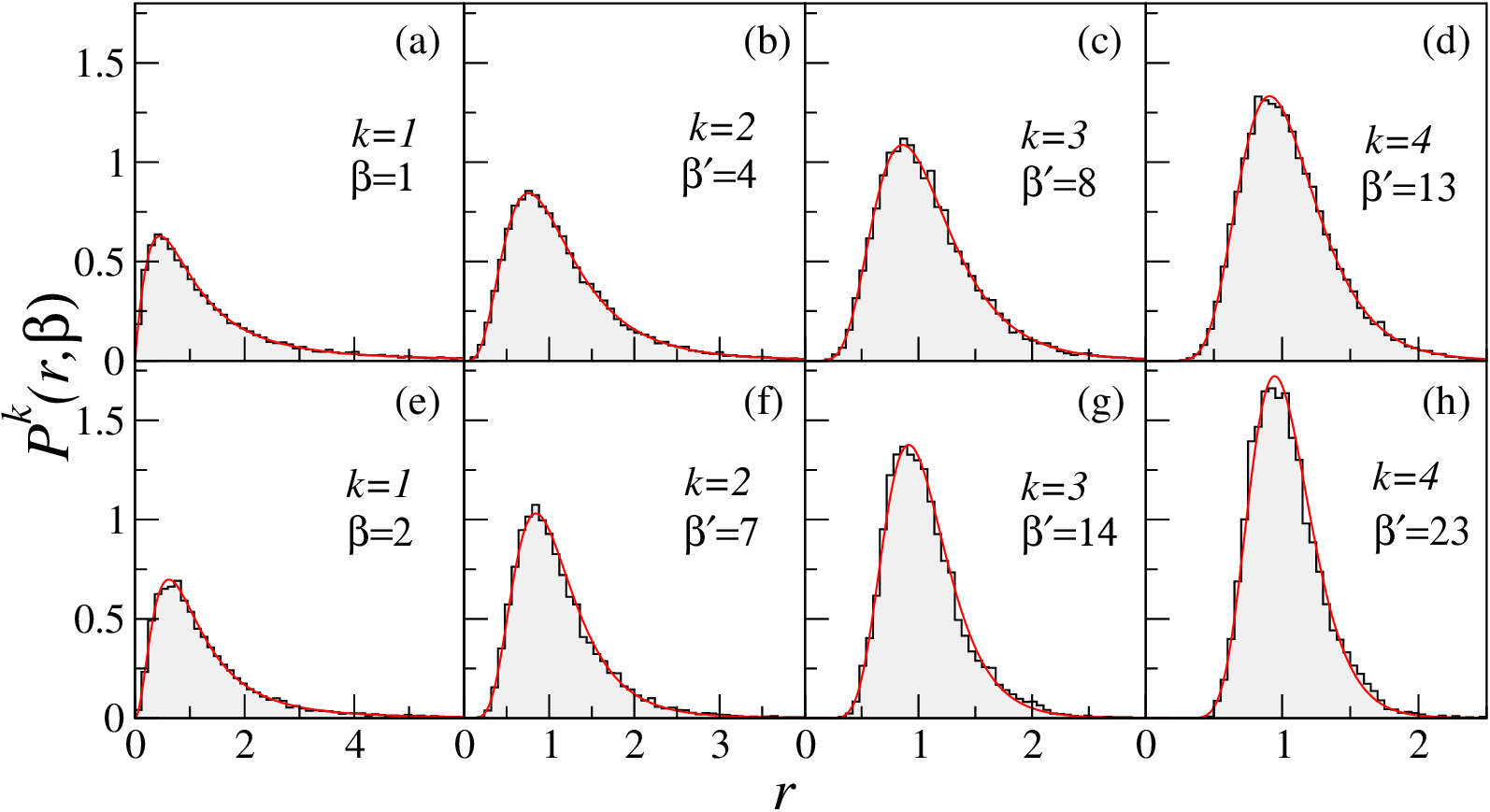}
\caption{The distribution of the $k$-th spacing ratios, for $k$=1, 2, 3, 4 is shown 
for Floquet systems; (upper panel) the kicked top, belonging to the COE class, and (lower panel) the 
intermediate map, belonging to the CUE class. The histograms are obtained from
computed eigenvalues of these systems, and the solid line represents $P(r,\beta')$, with 
$\beta'=$1, 4, 8, 13 for COE and $\beta'=$2, 7, 14, 23 for CUE.}
\label{coecue}
\end{figure}

The validity of Eq. \ref{eq2}-\ref{eq3} has also been
tested on time-dependent (driven) systems whose time-evolution operators (Floquet matrices)
have unimodular eigenvalues. Their fluctuation properties are modeled by those of
circular random matrix ensembles. One relevant model is the quantum kicked top whose
classical limit is chaotic \cite{kickedtop}. As this system is periodically kicked,
the quantum version can be studied in terms of the unitary time evolution operator
\begin{align}
\widehat{U}= \exp(-iqJ_z^2/2) ~ \exp(-ipJ_y),
\end{align}
where $q$=10 is the kick strength that acts as chaos parameter and $p=1.7$.
The action of this operator on a particle of angular momentum $\textbf{J}$, taken 
to be 200 here, is a precession about the $y$-axis, followed by state-dependent rotation 
about the $z$-axis as a consequence of periodic kicking.
The eigenvalues of $\widehat{U}$ are computed by diagonalizing this operator and 
its fluctuations are known to be consistent with COE statistics \cite{kickedtop}.
Figure \ref{coecue} (upper panel) shows the $k$-th spacing ratio distribution for this 
system which, as anticipated by Eqs. \ref{eq2}-\ref{eq3}, follows $P(r,\beta')$ with $\beta$=1.

As another instance of CUE class ($\beta=2$), a unitary operator corresponding to the so-called
intermediate map is considered. The quantum version of this map has been investigated previously
in the context of multifractal eigenstates, and in a specified range, has spectral fluctuations
similar to CUE matrices \cite{cuemap}. The unitary operator can be written in terms of 
an $N\times N$ matrix as
\begin{align}
 U_{ab}=\frac{\exp(-i\phi_a)}{N}\frac{1-\exp[i2\pi\gamma N]}{1-\exp[i2\pi(a-b+\gamma N)/N]},
\end{align}
with Hilbert space dimension $N=12000$. Here, $\phi_a$ is a random variable uniformly distributed
between [$0,2\pi$], and for any irrational $\gamma$ the spectral statistics is of the CUE type.
The computed distribution of $k$-th spacing ratios for this system, shown in Fig. \ref{coecue} (lower panel), 
agrees well with Eqs. \ref{eq2}-\ref{eq3}.

\section{Finite size effects}

\begin{figure}[t]
\includegraphics*[width=3.4in]{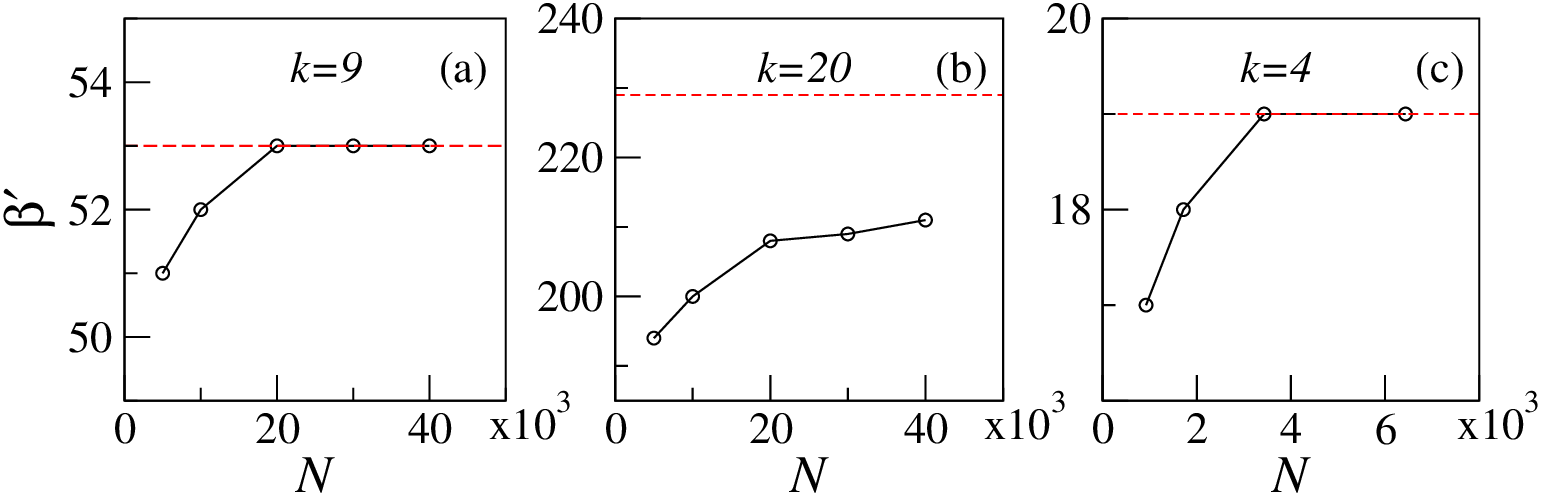}
\caption{Variation of $\beta'$ as a function of matrix dimension $N$, for random matrices of the GOE class,
for (a) $k=$9 and (b) $k=$20. For $k=9$, $\beta'$ converges to the predicted value ($\beta'=53$) as $N$ 
increases, while for $k=$20, a steady increase of $\beta'$ towards the predicted value of $\beta'=$229
is observed. (c) Variation of $\beta'$ as a function of matrix dimension $N$ for the GOE spin 
chain (Eq. \ref{schain1}). In this case, as $N$ increases, $\beta'$ converges to 19, the predicted value.}
\label{convergence}
\end{figure}

The scaling relation in Eqs. \ref{eq2}-\ref{eq3} suffers from finite size effects with
different systems converging to the scaling relation at different rates,
especially if $k>>1$. In the spectra of physical systems as well as in the random 
matrices of the Gaussian and circular ensembles, it was observed in practice that for higher 
order spacing ratios, say $k>$5, the value of $\beta'$ obtained by fitting $P(r,\beta')$
to the empirical distribution did not quite agree with that predicted by Eq. \ref{eq3}.
It is seen that the convergence to the predicted $\beta'$ is strongly pronounced 
as the order $N$ of the random matrix increases. This is illustrated
in Fig \ref{convergence} for two distinct values of $k$. In one case, for $k=9$ and based on
Eq. \ref{eq3}, the expected value of $\beta'=53$. Fig. \ref{convergence}(a)
shows a clear convergence to this predicted value as the order $N$ of the
random matrix increases. For $k=20$, Eq. \ref{eq3} predicts $\beta'$ to be 229.
However, as seen in Fig. \ref{convergence}(b), the convergence to the 
predicted value of $\beta'$ is rather slow, and up to $N$=40000 for which spectra was computed
it had not converged at all. Fig. \ref{convergence}(c) shows the same effect for the GOE
spin chain (Eq. \ref{schain1}) for $k=4$ as a function of the size of Hilbert space for 
the system. In this case, the dimension $N$ of the Hamiltonian matrix increases upon 
increasing the length $L$ of the spin chain. As Fig. \ref{convergence}(c) reveals convergence
is achieved for $N \approx 40000$.

In physical systems, it is well-known that purely quantum effects such as tunneling and
localization lead to deviations from random matrix averages.Hence, in physical systems, the deviations
from scaling could arise due to both physical effects that are not accounted for by RMT-type
universality and also the finite size effects. By studying these deviations in physical systems from 
expectations based on Eqs. \ref{eq2}-\ref{eq3} and comparing it with random matrices of identical
dimensions in which the deviations are purely due to finite size effects, it might be possible 
to distinguish whether the deviations occur due to finite size effects or system-dependent causes.
The distribution of higher-order ratios may then be useful to differentiate between and understand
the effects of random and system-dependent fluctuations in any physical system.

\section{Conclusions}
In summary, a generalized form of the Wigner surmise has been proposed to obtain
a form for the distribution of {\sl non-overlapping} spacing ratios of higher-orders, in which
the Dyson index $\beta \ge 4$ for the Gaussian and circular ensembles of random matrix theory.
An elegant scaling relation is observed, relating the order $k$ of the spacing ratio and
the Dyson index $\beta$, to another constant $\beta'$. Effectively, for large random matrix sizes 
$N>>1$, the form of distribution of higher-order spacing ratios $P^k(r,\beta')$ may be obtained 
for any arbitrary $k$, given the class of random matrices being considered.  For sufficiently small $N$
and large $k$, finite size effects do tend to induce deviations from the proposed scaling relation.
It has been shown that convergence to scaling relation can be restored by increasing the dimension
of the random matrix. In some cases, convergence to the predicted value was observed to be faster when the 
$k$-th order ratios considered were calculated in a completely non-overlapping manner, \textit{i.e.} in Eq. \ref{hosr},
$i$ increases in steps of $k$. Here, the numerator of the $i^{th}$ ratio, would be the denominator 
of the $(i+1)^{th}$ ratio.  The scaling relation Eq. \ref{eq3} remains unchanged when the distribution 
of these ratios are considered as well.

This scaling is shown to be valid for several different physical systems, namely, spin chains 
belonging to GOE and GUE class, measured resonances of Er atom, chaotic billiards, Floquet systems, 
all of whose eigenvalue fluctuations
properties are well described by an appropriate ensemble of random matrix theory.
It must be noted that higher-order spacing ratios are far easier to compute compared to the
spacing distribution beyond the nearest neighbor.
The results proposed here are of inherent interest in random matrix theory and provide
large number of additional statistic to easily test the fluctuation properties of physical systems 
for putative RMT-type behavior. The rigorous numerical results proposed in this work should
lead to attempts to obtain analytical justification for them.

\textit{Acknowledgements:} The authors acknowledge Sanku Paul for providing data for the 
levels of the intermediate map. UTB acknowledges the funding received from Department of Science 
and Technology, India under the scheme Science and Engineering Research Board (SERB) 
National Post Doctoral Fellowship (NPDF) file number PDF/2015/00050.

\end{document}